\documentclass[apj]{emulateapj}
\usepackage{graphicx}
\usepackage{amssymb}
\usepackage{color}

\newcommand{\be}{\begin{equation}}
\newcommand{\ee}{\end{equation}}
\newcommand{\ba}{\begin{eqnarray}}
\newcommand{\ea}{\end{eqnarray}}
\newcommand{\Msun}{M_\odot}
\newcommand{\Rsun}{R_\odot}
\newcommand{\Porb}{P_{\mathrm{orb}}}
\newcommand{\Porbdot}{\dot{P}_{\mathrm{orb}}}
\newcommand{\aK}{a}
\newcommand{\mx}{m_{\mathrm{x}}}
\newcommand{\mc}{m_{\mathrm{c}}}
\newcommand{\Rc}{R_{\mathrm{c}}}
\newcommand{\RL}{R_{\mathrm{L}}}
\newcommand{\mub}{B_8}
\newcommand{\Mdot}{\dot{M}}
\newcommand{\nus}{\nu_{\mathrm{s}}}
\newcommand{\nusdot}{\dot{\nu}_{\mathrm{s}}}
\newcommand{\nuseq}{\nus^{\mathrm{eq}}}
\newcommand{\mubeq}{B^{\mathrm{eq}}}
\newcommand{\mchat}{\hat{m}_{\mathrm{c}}}
\newcommand{\tdec}{\tau_{\mathrm{decay}}}
\newcommand{\tdecvz}{\tau_{\mathrm{decay}}^{\mathrm{VZ}}}
\newcommand{\tdeccs}{\tau_{\mathrm{decay}}^{\mathrm{CS}}}
\newcommand{\tsu}{\tau_{\mathrm{su}}}
\newcommand{\tsuvz}{\tau_{\mathrm{su}}^{\mathrm{VZ}}}
\newcommand{\tsucs}{\tau_{\mathrm{su}}^{\mathrm{CS}}}
\newcommand{\tsd}{\tau_{\mathrm{sd}}}
\newcommand{\nuone}{\nu_1}
\newcommand{\nutwo}{\nu_2}
\newcommand{\nuthree}{\nu_3}
\newcommand{\nuonevz}{\nuone^{\mathrm{VZ}}}
\newcommand{\nutwovz}{\nutwo^{\mathrm{VZ}}}
\newcommand{\nuthreevz}{\nuthree^{\mathrm{VZ}}}
\newcommand{\nuonecs}{\nuone^{\mathrm{CS}}}
\newcommand{\nutwocs}{\nutwo^{\mathrm{CS}}}
\newcommand{\nuthreecs}{\nuthree^{\mathrm{CS}}}

\shorttitle{Recycling of MSPs}
\shortauthors{Ho, Maccarone, \& Andersson}

\begin{document}
\title{Cosmic recycling of millisecond pulsars}
\author{Wynn C. G. Ho\altaffilmark{1}, Thomas J. Maccarone\altaffilmark{2},
 and Nils Andersson\altaffilmark{1}}
\altaffiltext{1}{School of Mathematics,
University of Southampton, Southampton, SO17 1BJ, United Kingdom;
wynnho@slac.stanford.edu, na@maths.soton.ac.uk.}
\altaffiltext{2}{School of Physics \& Astronomy,
University of Southampton, Southampton, SO17 1BJ, United Kingdom;
tjm@phys.soton.ac.uk.}

\begin{abstract}
We compare the rotation rate of neutron stars in low-mass X-ray binaries
(LMXBs) with the orbital period of the binaries.
We find that, while short orbital period LMXBs span a range of neutron
star rotation rates, all the long period LMXBs have fast rotators.
We also find that the rotation rates are highest for the systems with
the highest mean mass accretion rates, as can be expected if the
accretion rate correlates with the orbital period.
We show that these properties can be understood by a balance between
spin-up due to accretion and spin-down due to gravitational radiation.
Our scenario indicates that
the gravitational radiation emitted by these systems may be detectable by
future ground-based gravitational wave detectors.
\end{abstract}


\maketitle

\section{Introduction} \label{sec:intro}

Accreting millisecond X-ray pulsars are the evolutionary
link between the (million-year old) classical radio pulsars
and the (billion-year old) millisecond pulsars
\citep{smarrblandford76,alparetal82,radhakrishnansrinivasan82}.
The accreting millisecond pulsars are found in low-mass X-ray
binaries (LMXBs; \citealt*{chakrabartymorgan98,wijnandsvanderklis98}),
systems in which the pulsar is accreting from a low-mass stellar companion.
The details of how the transition from a rotating neutron star in
a low-mass X-ray binary to a millisecond radio pulsar takes place
are unknown \citep{bhattacharya95},
although a system has been found recently which shows evidence for having
made this transition \citep{archibaldetal09}.

We show here that a significant clue regarding the transition may be
contained in the relation between the rotation rate and orbital period
for the known accreting millisecond pulsars.
This trend suggests that a common physical mechanism(s) may be responsible.
We outline a simple scenario that can explain the evolution of the rotation
rate and orbital period for accreting millisecond pulsars and show how this
naturally leads to the observed population.
The proposed evolutionary scenario predicts gravitational wave emission
at a level that may be detected by next-generation ground-based
observatories.

\section{Data} \label{sec:data}

We use the tabulated values for the most rapidly-rotating ($\nus\ge 100$~Hz)
accreting millisecond pulsar (AMSP) spin frequency $\nus$,
orbital period $\Porb$, and mass accretion rate $\Mdot$
from \citet{galloway08,wattsetal08}\footnote{Note that
\citet{gallowayetal08,wattsetal08}
state that the orbital period of 4U~1702$-$429 is unknown.}.
The spin frequencies of some AMSPs are derived from brightness oscillations
seen during X-ray bursts \citep{strohmayeretal96,chakrabartyetal03},
which are due to thermonuclear burning of accreted material on the
surface of the neutron star \citep{strohmayerbildsten06}.
We do not consider spin frequencies derived from kHz quasi-periodic
oscillations since the correlation between the spin and quasi-periodic
oscillation frequencies is uncertain \citep{wijnandsetal03}.

\begin{figure}
\includegraphics[scale=0.4]{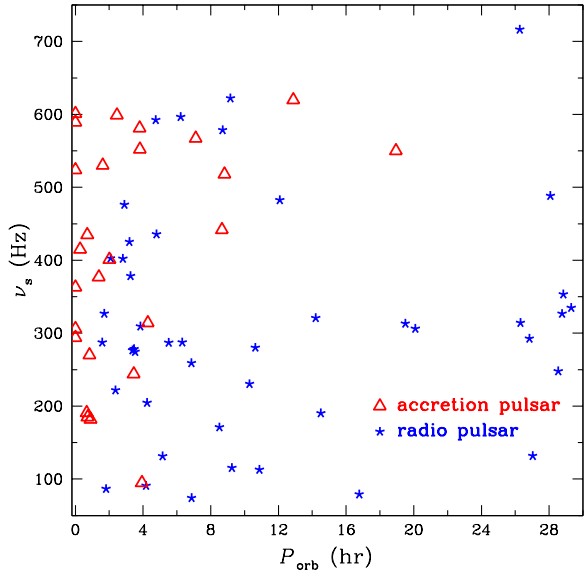}
\caption{
Observed spin frequency as a function of orbital period for the accreting
millisecond pulsars (triangles)
and rotation-powered radio pulsars (stars).
The spins and orbits for the accreting millisecond
pulsars are from \citet{galloway08,wattsetal08},
while those for the radio pulsars are from
the ATNF Pulsar Catalogue \citep{manchesteretal05}.
}
\label{fig:rpp}
\end{figure}

The trend in the measured values of the spin frequency and
orbital period for the AMSPs is shown in Fig.~\ref{fig:rpp}.
Also shown for comparison are values for the known (rotation-powered)
millisecond radio pulsars from the ATNF Pulsar
Catalogue\footnote{http://www.atnf.csiro.au/research/pulsar/psrcat}
\citep{manchesteretal05};
these rotation-powered pulsars are older and thought to be the
descendants of the AMSPs.
While the rotation-powered pulsars are distributed fairly uniformly
in the diagram of $\nus$ versus $\Porb$, the AMSPs are not.
In particular, at the highest rotation rates, the orbital periods of the AMSPs
span the range from about 0.5~hr to 20~hr, but there are no AMSPs with
$100\mbox{ Hz}\lesssim \nus< 440$~Hz and $\Porb>4.3$~hr.
A similar plot to Fig.~\ref{fig:rpp} was presented by \citet{kaaretetal06},
who remarked upon a possible absence of slowly-rotating, long orbital period
systems; they suggested that this may be due to an accretion rate that varies
with orbital period but made no further discussion.
A full population synthesis of neutron star spin and orbital periods
is beyond the scope of this work.
Nevertheless, a qualitative explanation of the $\nus$-$\Porb$ distribution of
AMSPs can be obtained by invoking simple physical processes that occur
in their evolution.

\section{AMSP Evolution} \label{sec:evol}

In close binary systems such as the LMXBs,
matter near the surface of the companion star is transferred to the
primary star (the latter is a neutron star in the case of AMSPs) when
the companion radius $\Rc$ is equal to the Roche lobe size $\RL$.
This size is the distance from the center of the donor star to the
inner Lagrangian point, where the gravitational forces from the companion
and primary are equal and opposite, and is given by \citep{paczynski71}
\be
\RL\approx 0.46\,\aK[\mc/(\mx+\mc)]^{1/3}, \label{eq:rochelobe}
\ee
where $\aK$ is the orbital separation between the neutron star of mass
$\mx$ and the companion star of mass $\mc$
($\mx$ and $\mc$ are in units of solar mass $\Msun$);
hereafter, we assume $\mx=1.4$.
When the companion is a main-sequence star,
$\mc\approx\Rc/\Rsun$, where $\Rsun$ is the solar
radius \citep{verbuntvandenheuvel95}.
Using Kepler's Third Law,
\be
\aK = 0.8\,\Rsun(\mx+\mc)^{1/3}(\Porb/\mbox{2 hr})^{2/3},
\ee
we obtain a relationship between the mass of the companion and the
orbital period \citep{kingetal96},
\be
\mc = 0.23\,\mchat(\Porb/\mbox{2 hr}), \label{eq:mcporb}
\ee
where $\mchat\equiv\mc/\mc^{\mathrm{MS}}$ and $\mc^{\mathrm{MS}}$ is the
mass of a main-sequence star that just fills its Roche lobe at $\Porb$;
the factor of $\mchat$~($<1$) is due to the fact that the companion star
in a binary has a larger radius for its mass than an isolated star
\citep[see, e.g.,][]{sillsetal00,andronovetal03}
and encapsulates our uncertainty in the (less evolved, since we only
consider $\mc<\mx$) evolutionary state of the companion.

We now examine the various physical processes that would lead to a
change in the angular momentum of the orbit or neutron star spin and
thus cause a source to move within the $\nus$-$\Porb$-plane of
Fig.~\ref{fig:rpp}.
Mass transfer from the companion to the neutron star can cause an
increase in the size of the orbit,
while orbital angular momentum loss due to magnetic braking or
gravitational wave emission causes the orbit to decrease.
The orbital period which separates expansion and decay is estimated to be
$>0.5$~days \citep[see, e.g.,][]{podsiadlowskietal02,mali09}.
For orbital periods longer than one hour, magnetic braking is dominant.
We consider two prescriptions for the magnetic braking torque.
The first one, from \citet{verbuntzwaan81}, results in a timescale for decay
\be
\tdecvz \approx (2.1\times 10^8\mbox{ yr})\,\mx^{2/3}\mchat^{-4/3}
 (\Porb/\mbox{2 hr})^{-2/3} \label{eq:tdecvz}
\ee
and a mass transfer rate $\Mdot_{-11}$
(in units of $10^{-11}\Msun\mbox{ yr$^{-1}$}$)
\citep[see also][]{kingetal96}
\be
\Mdot_{-11}^{\mathrm{VZ}}=80\,\mx^{-2/3}\mchat^{7/3}(\Porb/\mbox{2 hr})^{5/3}.
 \label{eq:mdotporbvz}
\ee
However, observations of rapidly rotating low-mass stars in open clusters
suggest that the rate of angular momentum loss described in
\citet{verbuntzwaan81} is too high (\citealt{sillsetal00,andronovetal03};
see also \citealt{yungelsonlasota08}, for more discussion).
Therefore we also use the magnetic braking torque from
\citet{chaboyeretal95,sillsetal00} \cite[see also][]{andronovetal03,mali09},
which results in
\ba
\tdeccs &\approx& (2.0\times 10^8\mbox{ yr})\,\mx^{2/3}\mchat
 (\Porb/\mbox{2 hr})^{7/3} \label{eq:tdeccs} \\
\Mdot_{-11}^{\mathrm{CS}}&=&87\,\mx^{-2/3}(\Porb/\mbox{2 hr})^{-4/3}.
 \label{eq:mdotporbcs}
\ea
Note that the mass accretion rate given by eq.~(\ref{eq:mdotporbvz})
increases with orbital period,
whereas that given by eq.~(\ref{eq:mdotporbcs}) decreases.
We also note that, at very short orbital periods $\Porb\lesssim 4$~hr,
gravitational radiation becomes important and some AMSP companions are
degenerate stars.
These systems may then be evolving to longer orbital periods, but they
will evolve slowly because the mass transfer rate and angular momentum
loss rate are low.
For example, SAX~J1808.4$-$3658 has a degenerate companion and increasing
orbital period with timescale $\Porb/\Porbdot\approx 6\times 10^7$~yr
\citep{podsiadlowskietal02,disalvoetal08,hartmanetal08,hartmanetal09}.

Accretion of angular momentum-carrying matter from the companion can
spin up the neutron star.  The timescale for spin-up is
$\tsu \approx (1.5\times 10^9\mbox{ yr})\,\mx^{-3/7}\mub^{-2/7}
\Mdot_{-11}^{-6/7}(\nus/\mbox{100 Hz})$,
where $\mub$ is the neutron star magnetic field
(in units of $10^{8}\mbox{ G}$).
Using the mass accretion rate from eqs.~(\ref{eq:mdotporbvz}) and
(\ref{eq:mdotporbcs}), we obtain
\ba
\tsuvz &=& (3.4\times 10^7\mbox{ yr})\,\mx^{1/7}\mub^{-2/7}\mchat^{-2}
 (\Porb/\mbox{2 hr})^{-10/7} \nonumber\\
&& \times (\nus/\mbox{100 Hz}) \nonumber\\
\tsucs &=& (3.2\times 10^7\mbox{ yr})\,\mx^{1/7}\mub^{-2/7}
 (\Porb/\mbox{2 hr})^{8/7} \nonumber\\
&& \times (\nus/\mbox{100 Hz}). \label{eq:tsu}
\ea
Spin-down of the neutron star by gravitational quadrupole radiation
occurs on a timescale
\be
\tsd \approx (2.9\times 10^{10}\mbox{ yr})\,\mx^{-2}\varepsilon_{-8}^{-2}
 (\nus/\mbox{100 Hz})^{-4}, \label{eq:tsd}
\ee
where $\varepsilon=10^{-8}\varepsilon_{-8}$ is the quadrupole
ellipticity.

We ignore spin-down by electromagnetic dipole radiation, which only becomes
dominant at $\mub\gtrsim 10\,\varepsilon_{-8}(\nus/100\mbox{ Hz})$.
It is instructive to estimate the magnetic field that would be required
for the AMSPs to be in magnetic spin equilibrium.
Magnetic spin equilibrium
occurs when the rotation period of the neutron star is equal to the
Keplerian period of the inner accretion disk at the magnetosphere
boundary \citep{davidsonostriker73};
this leads to an equilibrium spin period
\be
\nuseq=(270\mbox{ Hz})\,\mx^{5/7}\mub^{-6/7}\Mdot_{-11}^{3/7} \label{eq:nuseq}
\ee
or magnetic field
\be
\mubeq=(3.2\times 10^8\mbox{ G})\,\mx^{5/6}\Mdot_{-11}^{1/2}
 (\nus/\mbox{100 Hz})^{-7/6}.  \label{eq:magbeq}
\ee
Neutron stars with $\nus>\nuseq$ (or $B>\mubeq$) can act as mass-propellers,
as their rapid rotation creates a centrifugal barrier to accretion,
while stars with $\nus<\nuseq$ can accrete mass, gain angular momentum,
and be spun-up.

\begin{figure}
\includegraphics[scale=0.4]{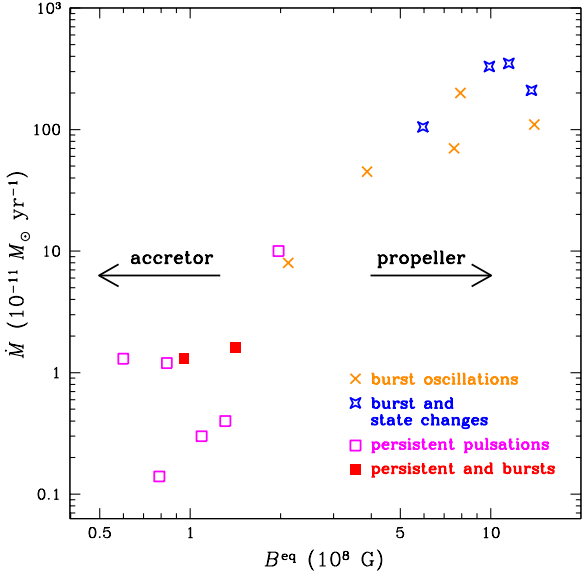}
\caption{
Magnetic field of accreting millisecond pulsars obtained from assuming the
neutron star rotates at spin equilibrium (see text)
for the observed mass accretion rate.
For a given $\Mdot$, neutron stars with $B>\mubeq$ are mass-propellers,
while stars with $B<\mubeq$ are accretors.
}
\label{fig:spineq}
\end{figure}

We consider whether the observed properties of neutron stars are
consistent with the idea that they are in magnetic spin equilbrium.
First, AMSPs are likely to have $B < \mubeq$
\citep{hartmanetal09,hartmanetal11,patruno10,papittoetal11}.
Using the maximum accretion rate observed during outbursts \citep{galloway08},
the high implied $\mubeq$ (see Fig.~\ref{fig:spineq})
could channel the accretion flow
and produce persistent coherent pulsations that are not detected;
the absence of persistent pulsations in the high accretion rate
systems, where the spin is generally measured from burst oscillations,
could be the result of magnetic screening by the accreted
material \citep{cummingetal01}.
In addition, some systems do not show changes to a propeller state at
the accretion rates estimated from spin equilibrium
\citep{barretolive02,maccaronecoppi03,gladstoneetal07}.

\begin{figure}
\includegraphics[scale=0.4]{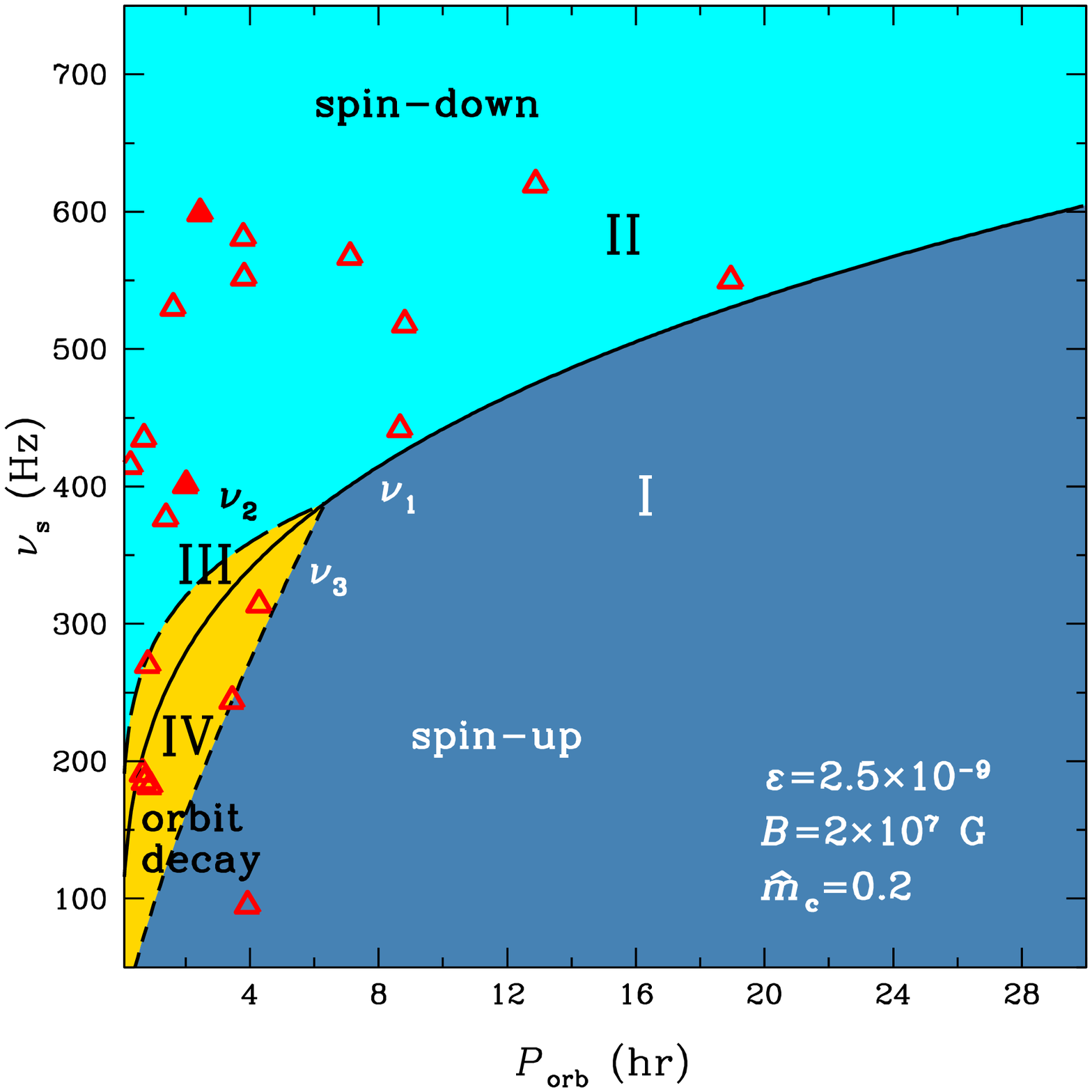}
\caption{
Evolution regimes of accreting millisecond pulsars:
AMSPs in the spin-up/spin-down regime (I/II) move to higher/lower spin
frequencies, as well as to shorter orbital periods but on a longer timescale;
AMSPs in the orbit decay regime (III/IV) move to shorter orbital
periods, as well as to lower/higher spin frequencies but on a longer timescale.
The lines (labeled $\nuonevz$, $\nutwovz$, $\nuthreevz$) separating the
regimes are given by eqs.~(\ref{eq:nuone})-(\ref{eq:nutwo}).
The open triangles denote AMSPs, while the solid triangles are the two AMSPs,
SAX~J1808.4$-$3658 at ($\Porb,\nus$)=(2~hr, 401~Hz) and
IGR~J00291$+$5934 at (2.5~hr, 599~Hz),
which show short-term spin-up and long-term spin-down (see text).
}
\label{fig:evolvz}
\end{figure}

\begin{figure}
\includegraphics[scale=0.4]{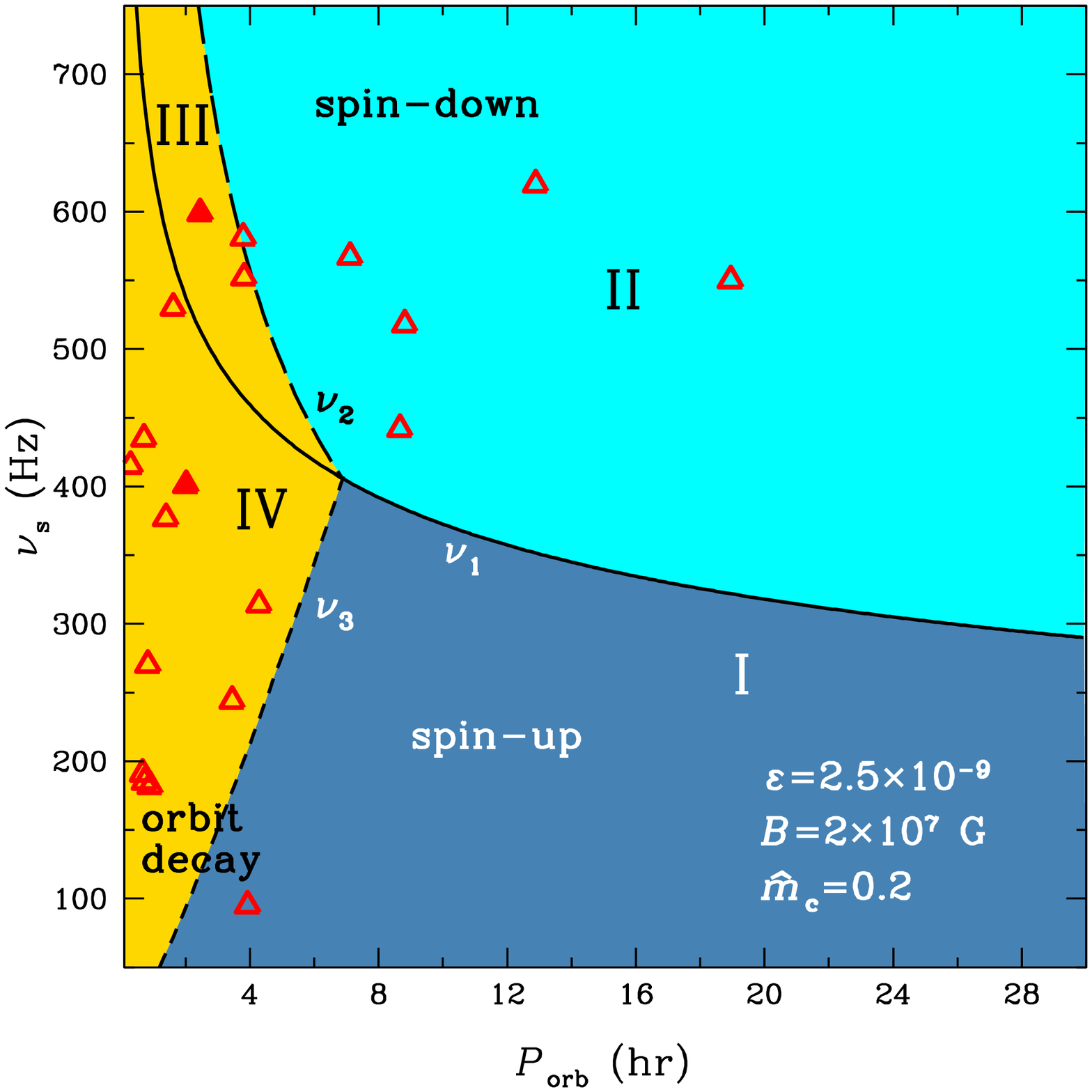}
\caption{
Evolution regimes of accreting millisecond pulsars:
AMSPs in the spin-up/spin-down regime (I/II) move to higher/lower spin
frequencies, as well as to shorter orbital periods but on a longer timescale;
AMSPs in the orbit decay regime (III/IV) move to shorter orbital
periods, as well as to lower/higher spin frequencies but on a longer timescale.
The lines (labeled $\nuonecs$, $\nutwocs$, $\nuthreecs$) separating the
regimes are given by eqs.~(\ref{eq:nuone})-(\ref{eq:nutwo}).
The open triangles denote AMSPs, while the solid triangles are the two AMSPs,
SAX~J1808.4$-$3658 at ($\Porb,\nus$)=(2~hr, 401~Hz) and
IGR~J00291$+$5934 at (2.5~hr, 599~Hz),
which show short-term spin-up and long-term spin-down (see text).
}
\label{fig:evolcs}
\end{figure}

The evolution of $\nus$ and $\Porb$ is primarily determined by the process
with the shortest timescale.
A comparison of the timescales is shown in Figs.~\ref{fig:evolvz}
and \ref{fig:evolcs}.
Over the ranges displayed, there are three regions/regimes:
spin-up from the accretion torque [see eq.~(\ref{eq:tsu})],
spin-down from gravitational quadrupole radiation [see eq.~(\ref{eq:tsd})],
and orbit decay from magnetic braking
[see eq.~(\ref{eq:tdecvz}) or (\ref{eq:tdeccs})].
An AMSP moves to higher $\nus$ (spin-up regime)
if the pulsar rotation rate is below both $\nuone$ (from $\tsu<\tsd$) and
$\nuthree$ (from $\tsu<\tdec$), where $\nuone$ is given by
\ba
\nuonevz &=& (\mbox{330 Hz})\,\varepsilon_{-8}^{-2/5}\mub^{2/35}\mchat^{2/5}
 (\Porb/\mbox{2 hr})^{2/7} \nonumber\\
\nuonecs &=& (\mbox{340 Hz})\,\varepsilon_{-8}^{-2/5}\mub^{2/35}
 (\Porb/\mbox{2 hr})^{-8/35} \label{eq:nuone}
\ea
and $\nuthree$ is given by
\ba
\nuthreevz &=&(\mbox{750 Hz})\,\mub^{2/7}\mchat^{2/3}(\Porb/\mbox{2 hr})^{16/21}
 \nonumber\\
\nuthreecs &=& (\mbox{740 Hz})\,\mub^{2/7}\mchat(\Porb/\mbox{2 hr})^{25/21}.
 \label{eq:nuthree}
\ea
If the rotation rate is above $\nuone$ and $\nutwo$ (from $\tsd<\tdec$),
where $\nutwo$ is given by
\ba
\nutwovz &=& (\mbox{270 Hz})\,\varepsilon_{-8}^{-1/2}\mchat^{1/3}
 (\Porb/\mbox{2 hr})^{1/6} \nonumber\\
\nutwocs &=& (\mbox{280 Hz})\,\varepsilon_{-8}^{-1/2}\mchat^{-1/4}
 (\Porb/\mbox{2 hr})^{-7/12}, \label{eq:nutwo}
\ea
then the AMSP moves to lower $\nus$ (spin-down regime).
Finally, a decrease in orbital period (orbit decay regime) occurs
when the rotation rate is below $\nutwo$ and above $\nuthree$.
We note that deviations from low values of $\mchat$
(up to $\sim 1$, e.g., due to degenerate companions; \citealt{kingetal96})
have the strongest effect on $\nuthree$, which do not change our results
qualitatively.

Though dependent on the various parameters,
the absolute timescale of the dominant process in each regime
bears out the observed population,
with a slight preference for a $\tsu$ that decreases with $\Porb$,
such as $\tsuvz$
(which is due to a magnetic braking torque that scales with $\Porb^\gamma$,
where $\gamma>1/3$).
In the spin-up region (long $\Porb$, low $\nus$),
\ba
\tsuvz &\sim& (7\times 10^6\mbox{ yr})\mchat^{-2}(\Porb/\mbox{10 hr})^{-10/7}
 (\nus/\mbox{200 Hz}) \nonumber\\
\tsucs &\sim& (4\times 10^8\mbox{ yr})(\Porb/\mbox{10 hr})^{8/7}
 (\nus/\mbox{200 Hz}).
\ea
In the spin-down region (high $\nus$), evolution is independent of orbital
period and occurs on a timescale
\be
\tsd\sim (2\times 10^9\mbox{ yr})(\varepsilon/10^{-9})^{-2}
(\nus/\mbox{500 Hz})^{-4}.
\ee
In the orbit decay region (short $\Porb$), evolution is independent
of spin frequency and occurs on a timescale
\ba
\tdecvz &\sim& (2\times 10^8\mbox{ yr})\mchat^{-4/3}(\Porb/\mbox{3 hr})^{-2/3}
 \nonumber\\
\tdeccs &\sim& (6\times 10^8\mbox{ yr})\mchat(\Porb/\mbox{3 hr})^{7/3}.
\ea
The two AMSPs, SAX~J1808.4$-$3658 \citep{hartmanetal09} and
IGR~J00291$+$5934 \citep{falangaetal05,patruno10,hartmanetal11,papittoetal11},
show an overall spin-down with timescales
$\nus/|\nusdot|\approx 2\times 10^{10}$~yr and $5\times 10^9$~yr, respectively
(interupted by short spin-ups during outbursts with timescales
$10^8$~yr and $2\times 10^7$~yr, respectively).

\section{Discussion} \label{sec:discuss}

We can now understand the observed population of AMSPs (especially the absence
of sources at low spin and long orbital period) as a result of the evolution
of $\nus$ and $\Porb$, depending on the process
(magnetic braking, mass accretion, and gravitational radiation)
with the shortest timescale.
AMSPs born at low $\nus$ and long $\Porb$ very quickly spin up to high
$\nus$.
Then on much longer timescales, these fast spinning sources slow down,
and their orbits decrease.
Once mass accretion ceases, there is no longer a spin-up torque;
the binary then contains a rotation-powered millisecond pulsar that can
move into the (former) spin-up region in Figs.~\ref{fig:evolvz}
or \ref{fig:evolcs} by spinning
down (as a result of gravitational wave or electromagnetic radiation)
or expanding its orbit (see Fig.~\ref{fig:rpp}).

We utilize two very different prescriptions for angular momentum loss
due to magnetic braking and find that we can vacate the low spin, long
orbital period spin-up region for reasonable parameter choices in both
cases.
We note a slight preference for the results obtained using
eqs.~(\ref{eq:tdecvz}) and (\ref{eq:mdotporbvz}),
as compared to eqs.~(\ref{eq:tdeccs}) and (\ref{eq:mdotporbcs}).
Future observations, for example, measurement of the mass accretion rate
dependence on orbital period, could distinguish between the two models.
There are presently six AMSPs with unknown $\Porb$, three of which have
$\nus<363$~Hz.
Measurement of a long orbital period for these systems or discovery of
new systems at the relatively low $\nus$
($100\mbox{ Hz}\lesssim\nus\lesssim 400\mbox{ Hz}$)
and long $\Porb$ would challenge the scenario proposed here.
Clearly more detailed studies of short orbital period systems
are required to address the trend seen,
as well as accounting for spin evolution in binary evolution models
\citep[see, e.g.,][]{podsiadlowskietal02,lambyu05,ferrariowickramasinghe07}.

The most rapidly-rotating millisecond radio pulsar \citep{hesselsetal06}
and AMSP have $\nus=716$~Hz and 620~Hz, respectively.
These frequencies are far below the theoretical maximum
(at $>1$~kHz), above which the centrifugal force causes
mass-shedding \citep{cooketal94,haenseletal99}.
It is thought that angular momentum loss from gravitational radiation
could be responsible for a spin limit below the mass-shedding
maximum \citep{wagoner84,bildsten98,anderssonetal99,melatospayne05}.
However, theoretical predictions for this limit are very uncertain,
with only the current gravitational wave searches \citep{abbottetal10}
and X-ray observations \citep{chakrabartyetal03} serving as constraints.
In our scenario, the observed position of AMSPs in Figs.~\ref{fig:evolvz}
and \ref{fig:evolcs} relative to the
different evolution regions
is suggestive that the amplitude of the mass quadrupole that produces
gravitational radiation is $\varepsilon\gtrsim 10^{-9}$.
At lower $\varepsilon$, spin-down by gravitational radiation
becomes irrelevant [see eqs.~(\ref{eq:nuone}) and (\ref{eq:nutwo})], and
all AMSPs would be in spin-up or orbit expansion or decay,
contrary to what is
seen \citep{hartmanetal09,hartmanetal11,patruno10,papittoetal11};
note that spin-down by electromagnetic radiation gives a similar region
in Figs.~\ref{fig:evolvz} and \ref{fig:evolcs} only if $B\gtrsim 10^9$~G.
Our estimated quadrupole ellipticity is far below the theoretical
maximum \citep{horowitzkadau09} of $4\times 10^{-4}$ and below
the current limit of $7\times 10^{-8}$ set by gravitational wave
detectors \citep{abbottetal10}.
Though observationally challenging \citep[see, e.g.,][]{wattsetal08},
future searches by Advanced LIGO or the proposed Einstein
Telescope \citep{punturoetal10,anderssonetal11} could provide direct
evidence for the evolutionary scenario outlined here.

\acknowledgements

The authors thank C.~O. Heinke, D.~I. Jones, and the anonymous referee
for comments that contributed to improvements to the manuscript.
WCGH appreciates the use of the computer facilities at the Kavli
Institute for Particle Astrophysics and Cosmology.
WCGH and NA acknowledge support from the Science and
Technology Facilities Council (STFC) in the United Kingdom.

\bibliographystyle{apj}

\end{document}